\documentclass[prl,aps,twocolumn,footinbib,showpacs,preprintnumbers,superscriptaddress]{revtex4-1}
\usepackage{graphicx}% Include figure files
\usepackage{dcolumn}% Align table columns on decimal point
\usepackage{bm}% bold math
\usepackage{amssymb}
\usepackage{pifont}
\usepackage{amsmath}
\usepackage{times}

\usepackage{graphicx,psfrag}

\usepackage{xcolor} 
 
\begin{document}

\title{Kuramoto model for excitation-inhibition-based oscillations}
\author{Ernest Montbri\'o}
\affiliation{Center for Brain and Cognition. Department of Information and Communication Technologies,
Universitat Pompeu Fabra, 08018 Barcelona, Spain}
\author{Diego Paz\'o}
\affiliation{Instituto de F\'{i}sica de Cantabria (IFCA), CSIC-Universidad de
Cantabria, 39005 Santander, Spain }
 
\date{\today}
%%%%%%%%%%%%%%%%%%%%%%%%%%%%%%%%%%%%%%%%%%%%%%%%%%%%%%%%%%%%%%%%
 
\begin{abstract}
The Kuramoto model (KM) is a theoretical paradigm for investigating  
the emergence of rhythmic activity in large populations of oscillators.
A remarkable example of rhythmogenesis is the feedback loop between 
excitatory (E) and inhibitory (I) cells in large neuronal networks. Yet, although 
the EI-feedback mechanism plays a 
central role in the generation of brain oscillations, it remains unexplored  
whether the KM has enough biological realism to describe it. 
Here we derive a two-population KM
that fully accounts for the onset of EI-based neuronal rhythms and
that, as the original KM, is analytically solvable to a large extent.
Our results provide a powerful theoretical tool for
the analysis of large-scale neuronal oscillations.
\end{abstract}
 
\maketitle

The Kuramoto model (KM) is an idealized mathematical model 
for exploring the birth of collective synchronization in its most 
simple form. It consists of a population of heterogeneous, 
all-to-all coupled oscillators, and is a unique
example of exactly solvable system of nonlinear differential equations
~\cite{Kur84,Str00,PRK01,ABP+05,PR15i}. 
Yet, the KM was originally not intended as a specific    
description of any particular system,
and finds limited applications in the modeling and analysis of
natural oscillatory phenomena, see e.g.~\cite{WCS96,SAM+05,KZH02}.

An important example of collective synchronization are
large scale neuronal oscillations
~\cite{Buz06,PRK01}. Despite continued work using the KM to investigate 
neuronal rhythms
(see e.g.~\cite{BHD10,CHS+11,TDD14,VMM14,PDH+15,SKS+15,PSP+16}),
it remains unknown whether the KM actually accounts for
the neuronal mechanisms resulting in such oscillations.
In this Letter we derive a simple, two-population KM,
that describes one of the basic  mechanisms of
generation of neuronal oscillations: 
The feedback loop between fast excitation (E)
and slow inhibition (I)
in large neuronal networks~\cite{WTK+00,TS09,Wan10,BW12}.

\paragraph{EI-feedback loop and EI-based oscillations:}
The canonical neuronal network to model the EI-feedback loop consists of 
two interacting populations of
excitatory and inhibitory  neurons~\cite{WC72,BW03,HM03,BK03}.
Here, we consider two populations of 
$N$ pulse-coupled ``Winfree oscillators''
~\cite{Str00,Win67,AS01,PM14,GMP17,Lai14i} 
with phase variables $\{\theta^\sigma_i\}_{i=1,\ldots,N}$ 
(populations are identified by $\sigma \in \{E,I\}$), 
which evolve according to
\begin{equation}
\dot\theta_i^\sigma =\omega_i^\sigma +\xi_i^\sigma + Q(\theta^\sigma_i) \, 
 \left(K_{\sigma E}  h_E - K_{\sigma I} h_I \right) .
\label{winfree}
\end{equation}
The natural frequencies
$\omega_i^{\sigma}$ are drawn from  
Lorentzian distributions of half-width $\gamma$,
centered at $\bar \omega_\sigma$
\begin{equation}
g_\sigma(\omega)=(\gamma/\pi) \left[(\omega -\bar\omega_\sigma)^2+\gamma^2\right]^{-1},
\label{lorentzian}
\end{equation}
and $\xi_i^\sigma$ are independent, zero-mean delta-correlated noise
processes of strength $D$:  $\langle \xi_i^\sigma(t) \, \xi_j^{\sigma'}(t')\rangle=  2 D \delta(t-t') \delta_{i,j} \delta_{\sigma,\sigma'}$.
In Eq.~\eqref{winfree}, $Q(\theta)$ is the so-called phase response curve (PRC)
that determines the response of the oscillators to perturbations.
Here we adopt the (infinitesimal) PRC of the theta-neuron model, 
$Q(\theta)= 1-\cos \theta$, which is nonnegative and thus results in
phase advances/delays in response to excitatory/inhibitory inputs~\cite{EK86,Erm96,Izh07}. 
Neuronal oscillators with nonnegative PRC are called Type 1, 
and include a broad class of neuronal models, see e.g.~\cite{Erm96,Izh07,BHH04}.
The oscillators interact all-to-all via the mean fields 
\begin{equation}
h_{\sigma} = \frac{1}{N} \sum_{j=1}^N P(\theta^\sigma_j) ,
\label{h0}
\end{equation}
which are population-averaged sums of all the pulses $P$ produced in each population.
We use  
the family of unimodal even-symmetric functions
$P(\theta)= (1-r)(1+\cos\theta)(1-2r\cos\theta +r^2)^{-1}$, 
with $\int_{-\pi}^{\pi} P(\theta) d\theta= 2\pi$ and a free parameter $r\in(-1,1)$, 
such that $\lim_{r\to1}P(\theta)=2\pi\delta(\theta)$~\cite{GMP17}.
Expressed in words, the
$j$-th oscillator in the $E$ population exerts a positive, 
pulse-like influence $P(\theta^E_j)$ of strength $K_{E E}/N\geq 0$ 
to each oscillator of
the $E$ population, and of strength 
$K_{IE}/N\geq 0$ to each oscillator
of the $I$ population 
(similarly for the $j$-th oscillator of the $I$ population, with 
an explicit ``$-$'' sign in Eq.~\eqref{winfree} corresponding to inhibition).

%%%%%%%%%%%%%%%%%%%%%%%%%%%%%%%%%%%%%%%%%%%%%%%%%%%%%%%%%%%%%%%%%%%%%%%%
\begin{figure}[b]
\psfrag{Om/D}[b][b][1.]{$\Omega/\Delta$}
\psfrag{J/D}[b][b][1.]{$J/\Delta$}
\psfrag{Om=0}[b][b][1.]{$\tilde \Omega=0$}
\psfrag{e=0}[b][b][1.]{$\epsilon=0$}
\psfrag{t}[b][b][1.]{$t$}
\psfrag{P}[b][b][.9]{$h_E,h_I$}
\psfrag{n}[b][b][.75]{Oscillator number}
\psfrag{a}[b][b][.9]{{\bf(a)}}
\psfrag{b}[b][b][.9]{{\bf(b)}}
\psfrag{c}[b][b][.9]{{\bf(c)}}
\psfrag{d}[b][b][.9]{{\bf(d)}}
\centerline{\includegraphics[width=75mm,clip=true]{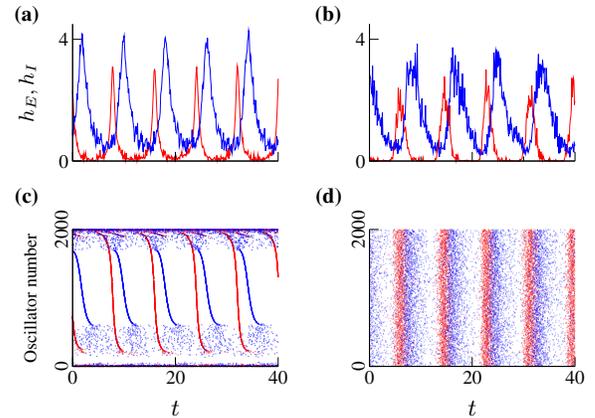}}
\caption{EI-based oscillations in a population of $N=2000$ excitatory (E) 
and $N=2000$ inhibitory (I) Winfree oscillators, Eq.~\eqref{winfree};
with $\bar\omega_E=1.5$, $\bar\omega_I=0.5$, $K_{EI}=K_{IE}=0.5$, 
$K_{EE}=K_{II}=0$, and $r=0.99$.
(a,b) Time series of the  E (red) and I (blue) activity-based mean fields
$h_{\sigma}$. (c,d) Raster plots: 
A point is plotted when an oscillator's phase reaches a multiple of $2\pi$, 
which is the peak location of $P(\theta)$.
In (a,c) frequencies are Lorentzian distributed,
with $\gamma=0.1$, and $D=0$. In (b,d) the noise strength is $D=0.1$, and $\gamma=0$.
}
\label{Fig1}
\end{figure}
%%%%%%%%%%%%%%%%%%%%%%%%%%%%%%%%%%%%%%%%%%%%%%%%%%%%%%%%%%%%%%%%%%%%%%%%

Figure \ref{Fig1}(a,b) shows EI-based oscillations
of the mean-field quantities $h_\sigma$ 
in simulations of (a) heterogeneous and (b) 
noisy EI-Winfree networks, Eqs~\eqref{winfree}.
The raster plots Figs.\ref{Fig1}(c,d) show that an EI-oscillation cycle
begins with the synchronous `firing' of  
a large cluster of phase-locked
E-oscillators, followed by another synchronous `firing'
of the I-oscillators. 
Note that, to emphasize that oscillations emerge exclusively 
due to the interplay between fast excitatory and slow inhibitory dynamics,
in Fig.~\eqref{Fig1} 
we set the self-coupling terms to zero, $K_{EE}=K_{II}=0$,
and consider $\Delta \omega \equiv \bar \omega_E-\bar \omega_I>0$.
In the following we derive a two-population KM that captures the main features of
the oscillations shown in Fig.~\eqref{Fig1}, and that is exactly solvable to 
a large extent.

\begin{figure}[t]
\psfrag{R}[b][b][.9]{$R_E,R_I$}
\psfrag{t}[b][b][1.]{$t$}
\psfrag{h}[b][b][.9]{$h_E,h_I$}
\psfrag{Oscillator number}[b][b][.75]{Osc. number}
\psfrag{a}[b][b][.9]{{\bf(a)}}
\psfrag{b}[b][b][.9]{{\bf(b)}}
\psfrag{c}[b][b][.9]{{\bf(c)}}
\centerline{\includegraphics[width=70mm,clip=true]{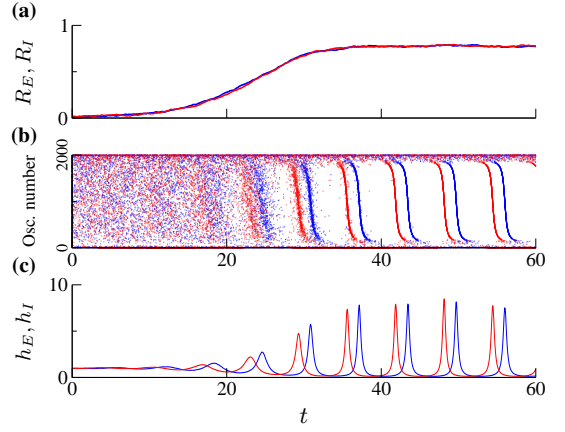}}
\caption{EI-based oscillations in the EI-KM Eq.~\eqref{EI-KM} 
with quenched heterogeneity  and $N=2000$. 
(a) Amplitude of the Kuramoto order parameters, $R_E$ (red) and $R_I$ (blue);  (b) Raster plots;
 (c) Mean fields obtained applying Eq.~\eqref{h} to $Z_\sigma$; 
Parameters are as in Fig.~\ref{Fig1}(a,c), 
except that here $r=1$, instead of $r=0.99$.}
\label{Fig2}
\end{figure}

\paragraph{Excitation-Inhibition Kuramoto model (EI-KM):}
Invoking the averaging approximation, valid for weak coupling and nearly 
identical oscillators~\cite{Kur84,PRK01},  the EI-Winfree model in
Eq.~\eqref{winfree} reduces to the EI-KM
\footnote{  Phase variables in
Eqs.~\eqref{EI-KM} correspond to slow-phase approximations 
of the phases in Eqs.~\eqref{winfree}. 
See Supplemental Material, Section~I.A.}
\begin{eqnarray}
 \label{EI-KM} 
\dot \theta_i^\sigma &=& \tilde \omega_i^\sigma 
 +\xi_i^\sigma \\
&-&\frac{1+r}{2N}  \sum_{j=1}^N \left[ 
 K_{\sigma E} \cos ( \theta_i^\sigma-\theta_j^E )
- K_{\sigma I}  \cos (\theta_i^\sigma- \theta_j^I) 
  \right ] , \nonumber
\end{eqnarray}
where $\tilde \omega_i^\sigma \equiv \omega_i^\sigma +K_{\sigma E} - K_{\sigma I}$.
There are two major differences between the EI-KM and the classical 
single and two-population KM broadly investigated in the 
literature, see e.g.~\cite{Kur84,OK91,MKB04,BHO+08,AMS+08,KNA+10a,PDD16}.
First, in the EI-KM the excitatory and inhibitory coupling constants 
differentially shift the natural frequencies $\tilde \omega_i^E$ 
and $\tilde \omega_i^I$, and this largely affects the regions 
of parameters where EI-oscillations occur. 
Second, 
although the cosine coupling does not promote synchrony 
in the KM~\cite{SK86}, the positive (E) and negative (I) cross-coupling terms in 
Eqs.~\eqref{EI-KM} crucially conspire to synchronize the oscillators
%%%%%%%%%%%%%%%%%%%%%%%%%%%%%%%%%%%%%%%%%%%%%%%%%%%%%%%%%%%%%%%
\footnote{
Using $\phi^E_i \equiv \theta^E_i-\pi/2$, Eqs.~\eqref{EI-KM} transform into a 
two-population model with phases $\{\phi_i^E\}$, $\{\theta_i^I\}$, 
where self-interaction functions remain the same,
while cross-interaction functions become sine functions
with the precise signs to favor synchrony.
Hence, for the case $K_{EE}=K_{II}=0$ considered in Fig.~\ref{Fig2}, the system 
reduces to a bipartite network of Kuramoto oscillators.
}.
%%%%%%%%%%%%%%%%%%%%%%%%%%%%%%%%%%%%%%%%%%%%%%%%%%%%%%%%%%%%% 
Therefore, in the EI-KM synchrony sets in exclusively 
due to the cooperative action of both the E and the I populations, 
in consonance with the EI-feedback loop mechanism.
Indeed, Fig.~\ref{Fig2} shows numerical simulations of the 
EI-KM in Eqs.~\eqref{EI-KM} using the same parameters as in 
Fig.~\ref{Fig1}(a,c) ---except $r$,
which in the EI-KM is set to $r=1$, see below.
Fig.~\ref{Fig2}(a) displays the amplitude of the 
complex Kuramoto order parameters $Z_\sigma\equiv R_{\sigma} e^{i\Psi_{\sigma}}=
N^{-1} \sum_{j=1}^{N}e^{i\theta_j^{\sigma}}$. 
At $t=0$, the amplitudes $R_{E}$ and $R_{I}$ are near zero since the 
initial values of all the phases are randomly distributed in the interval
$[0,2\pi)$. Then, after a brief transient, 
the Kuramoto order parameters converge (up to finite-size fluctuations) 
to uniformly rotating solutions $Z_\sigma(t)=R_* e^{i \Psi_\sigma(t)}$, with $0 <R_*<1$
 and $\dot\Psi_\sigma=\Omega$, signaling the onset of collective synchronization.
Note that the raster plot in Fig.~\ref{Fig2}(b) shows  
that the cluster of E oscillators precedes the cluster of I oscillators, 
consistent with Fig.~\ref{Fig1}(c).

Finally, in the EI-KM the width of the pulses (controlled by $r$) 
influences the intensity of the cosine coupling functions.
To lighten the notation, hereafter we set $r=1$ in Eqs.~\eqref{EI-KM},
corresponding to the limit of infinitely narrow (Dirac delta) pulses 
---this is close to the value used in Fig.~\eqref{Fig1}. The 
generalization of our results to general $r$ is trivial.

%%%%%%%%%%%%%%%%%%%%%%%%%%%%%%%%%%%%%%%%%%%%%%%%%%%%%%%%%%%%%%%%%%%%%%%%
%%%%%%%%%%%%%%%%%%%%%%%%%%%%%%%%%%%%%%%%%%%%%%%%%%%%%%%%%%%%%%%%%%%%%%%

\begin{figure}%[b]
\psfrag{Om/D}[b][b][1.]{$\Delta \omega/\gamma$}
\psfrag{J/D}[b][b][1.]{$K/\gamma$}
\psfrag{Om=0}[b][b][1.]{$\Delta \omega/\gamma=0$}
\psfrag{e=0}[b][b][1.]{$\epsilon=0$}
\psfrag{e=1}[b][b][1.]{$\epsilon=1$}
\psfrag{e=2}[b][b][1.]{$\epsilon=2$}
\psfrag{e=3}[b][b][1.]{$\epsilon=3$}
\psfrag{a}[b][b][.9]{{\bf(a)}}
\psfrag{b}[b][b][.9]{{\bf(b)}}
\psfrag{c}[b][b][.9]{{\bf(c)}}
\psfrag{d}[b][b][.9]{{\bf(d)}}
\centerline{\includegraphics[width=75mm,clip=true]{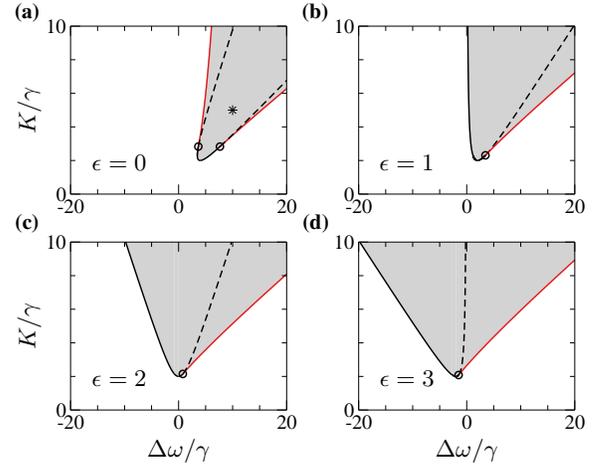}}
\caption{Phase diagrams of the EI-KM~Eq.~\eqref{EI-KM}
with $D=0$ and coupling constants given by Eq.~\eqref{K}, for 
various values of $\epsilon$. 
Regions of stable synchronization are highlighted in gray. 
Synchronization and incoherence are both stable 
in regions limited by black-dashed and red lines.  
The asterisk in (a) marks the parameter values used in 
Fig.~\ref{Fig2}.
Black lines correspond to Eq.~\eqref{Hopf}. Solid and dashed lines 
are separated by codimension-2 points ---obtained from Eq.~\eqref{c2i}---,
and indicate super/sub-critical 
bifurcations, respectively. 
Red curves indicate saddle-node bifurcations.}
\label{Fig3}
\end{figure}

\paragraph{Analysis of the EI-KM:} 
Eqs.~\eqref{EI-KM} can be 
efficiently analyzed in the thermodynamic limit, 
$N\to\infty$. 
To do so, the discrete sets of phases and frequencies 
turn into continuous variables 
$\{\theta_i^{\sigma},\omega_i^{\sigma}\}\to \{\theta_\sigma,\omega_\sigma\}$, 
and the corresponding probability 
density functions $f^{\sigma}(\theta_\sigma|\omega_\sigma,t)$ 
satisfy 
coupled the Fokker-Planck equations 
\begin{equation}
\partial_t f^\sigma= -\partial_{\theta_\sigma}\left(f^\sigma \dot \theta_\sigma\right) + D \,  \partial_{\theta_\sigma}^2 f^\sigma, 
\label{fk}
\end{equation}
for which the fully incoherent state 
$f^E=f^I=(2\pi)^{-1}$ is always a trivial solution \cite{OK91,SM91}. 
It is convenient to introduce the Fourier expansion of $f^{\sigma}$:
\begin{equation}
f^\sigma (\theta|\omega,t)= \frac{1}{2\pi}\sum_{l=-\infty}^{\infty} 
f_l^\sigma(\omega,t) e^{il\theta}, 
\label{fourier}
\end{equation}
where $f_0^\sigma=1$ and $(f^\sigma_{-l})^*=f^\sigma_{l}$ 
(the asterisk denotes complex conjugate).    
Thus, the Kuramoto order parameters are 
\begin{equation}
Z_\sigma = \left[ \int_{-\infty}^\infty 
f^{\sigma}_{1} (\omega,t) \,  g_\sigma(\omega) \, d\omega \right]^*.
\label{op}
\end{equation}
Substituting Eq.~\eqref{fourier} into Eq.~\eqref{fk}, 
yields two infinite sets of integro-differential equations 
for the Fourier modes
\begin{eqnarray}
\dot f^\sigma_l &=&  - (i l  \tilde \omega_\sigma+l^2 D) f^\sigma_l+ \frac{i l}{2}  f^\sigma_{l-1}  
( K_{\sigma E} Z_E^* - K_{\sigma I}Z_{I}^*)  \nonumber \\
&& +\frac{i l}{2} f^\sigma_{l+1} ( K_{\sigma E} Z_E- K_{\sigma I}Z_I) ,
\label{modes} 
\end{eqnarray}
where $\tilde\omega_\sigma \equiv \omega_\sigma+K_{\sigma E}-K_{\sigma I}$.
The stability of the incoherent state can be analyzed by 
linearizing Eq.~\eqref{modes}
\footnote{See the Supplemental Material, Section~II.}. 
To simplify the analysis, we study the case in which  
cross- and self-couplings are symmetric, 
\begin{equation}
 K_{EI} =K_{IE} \equiv K , \quad K_{II}=K_{EE} \equiv \epsilon K,
\label{K}
\end{equation}
and use the new parameter $\epsilon \geq 0$ as a measure of the ratio of self- to 
cross-coupling. 
Then we find that the eigenvalues determining the stability of incoherence are
\begin{equation}
\lambda_\pm= -\gamma-D \pm \tfrac{1}{2}\sqrt{K^2-[\Delta\omega+(\epsilon-2)K]^2}-i\Omega, 
\label{lambda}
\end{equation}
where $\Omega = (\bar\omega_E+\bar\omega_I)/2$ is  
the center of the frequency distribution combining E and I populations.
Note that parameters $\gamma$ and $D$ play identical roles in
Eq.~\eqref{lambda}, as it occurs in the KM~\cite{SM91}. 
Imposing Re$(\lambda_+)=0$ in Eq.~\eqref{lambda}, 
we find the boundary of incoherence 
\begin{equation}
\left( \frac{\Delta\omega}{\gamma+D}\right)_{c}^{\pm}=  (2-\epsilon) 
\frac{K}{\gamma+D} \pm \sqrt{\left ( \frac{K}{\gamma+D}\right)^2-4},
\label{Hopf}
\end{equation}  
which is the family of hyperbolas depicted by solid and dashed
black lines
in Figs.~\ref{Fig3}(a-d), for increasing values of $\epsilon$. 
A necessary condition for the boundary Eq.~\eqref{Hopf} to exist is 
\begin{equation}
\frac{K}{ \gamma+D} \geq 2.
\label{cond}
\end{equation}  
Hence, given a certain level of
heterogeneity and/or noise, synchronization sets in at large enough 
values of the coupling strength. This is remarkably similar 
to the KM~\cite{Kur84,SM91}, 
although here $K$ represents cross-, and not self-coupling. 
Moreover,
Eq.~\eqref{cond} is not a sufficient condition for synchronization 
in the EI-KM. If Eq.~\eqref{cond} is satisfied, 
then Eq.~\eqref{Hopf} shows that synchronization is only achieved 
for a particular range of values of the frequency mismatch $\Delta \omega$.  
The coupling ratio $\epsilon$ does not affect Eq.~\eqref{cond}, 
but it critically controls the range of $\Delta \omega$ 
for stable incoherence:
Note that when $\epsilon\le1$, the boundary Eq.~\eqref{Hopf} is 
located at positive values of $\Delta \omega$, and   
thus incoherence is always stable when
 I oscillators are intrinsically faster than E
oscillators ($\Delta\omega<0$), see Fig.~\eqref{Fig3}. 
Increasing the parameter $\epsilon$
shifts the boundary, with asymptotes at $K=\Delta\omega/(3-\epsilon)$ 
and $K=\Delta\omega/(1-\epsilon)$,
towards negative values of $\Delta \omega$. Thus, increasing the 
coupling ratio through $\epsilon$ provides a key ingredient for 
synchronizing EI networks when $\bar \omega_I>\bar \omega_E$,
as I-to-I coupling slows down I oscillators while  
E-to-E coupling speeds up  E oscillators.
 
 %%%%%%%%%%%%%%%%%%%%%%%%%%%%%%%%%%%%%%%%%%%%%%%%%%%%%%%%%%%%%%%%%%%%%%%%
%%%%%%%%%%%%%%%%%%%%%%%%%%%%%%%%%%%%%%%%%%%%%%%%%%%%%%%%%%%%%%%%%%%%%%%

The synchronization region turns out to be larger than the hyperbolic
boundary defined by Eq.~\eqref{Hopf}, particularly for large $\epsilon$ values 
(see Fig.~\ref{Fig3} for the noise-free case). 
The reason is that the bifurcation at Eq.~\eqref{Hopf}
is often sub-critical. 
To investigate this further,
next we consider the purely heterogeneous ($D=0$) 
and the purely noisy ($\gamma=0$) cases separately, and show that 
the global picture is remarkably similar in both  instances.

%%%%%%%%%%%%%%%%%%%%%%%%%%%%%%%%%%%%%%%%%%%%%%%%%%%%%%%%%%%%%%%%%%
The noise-free problem is particularly simple
since it can be assumed 
that the densities in Eq.~\eqref{fourier} satisfy the so-called 
Ott-Antonsen (OA) ansatz~\cite{OA08,OA09}
\begin{equation}
f^\sigma_{l>1} (\omega,t)= [f^\sigma_1 (\omega,t)]^l.
\label{OAA}
\end{equation}
A first useful outcome of the OA ansatz is that it allows to 
infer the mean field $h_\sigma$, Eq.~\eqref{h0}, 
from the Kuramoto order parameter $Z_\sigma$, Eq.~\eqref{op}.   
Specifically, in the thermodynamic limit 
$h_\sigma(t)=\int_{-\infty}^{\infty} \int_{0}^{2\pi} 
P(\theta) f^\sigma(\theta|\omega,t) g_\sigma(\omega) d\omega d\theta$. Then,
considering   
$P(\theta)$ as defined above,  and the heterogeneity in
Eq.~\eqref{lorentzian}, one finds   
$h_\sigma=\mathrm{Re} [ (1+Z_\sigma)/(1-rZ_\sigma)]$, 
see~\footnote{See Supplemental Material, Section~III.B and~\cite{GMP17}.}.
In the limit $r\to 1$, this relation reduces to  
\begin{equation}
h_\sigma= (1-R_\sigma^2)(1+R_\sigma^2-2R_\sigma \cos \Psi_\sigma)^{-1}. 
\label{h}
\end{equation}
Figure~\ref{Fig2}(c) displays the mean fields 
$h_\sigma(t)$ obtained applying Eq.~\eqref{h} 
to the Kuramoto order parameters $Z_\sigma(t)$ of the EI-KM.
It can be seen that uniformly rotating solutions of  
the Kuramoto order parameters correspond to 
pulsatile oscillations of the 
activity-based mean fields $h_\sigma(t)$ 
\footnote{
In Section~III.A of the Supplemental Material we show that, in the EI-KM, the mean field $h_\sigma$ 
is linearly related with the \emph{mean firing rate} 
of the population of oscillators that, compared to $h_\sigma$, 
is a more natural measure of neuronal 
activity in neuroscience.
}.
Though the agreement between Figs.~\ref{Fig1}(a) and \ref{Fig2}(c) is only 
qualitative, it gradually improves as parameters  $\gamma$ and $\Delta\omega$ 
are decreased and the averaging approximation becomes more accurate  
~\footnote{
See Supplemental Material, Section~I.B.
}. 

\begin{figure}
\psfrag{Psi}[b][b][1.]{$\Phi_*$}
\psfrag{R}[b][b][1.]{$R_*$}
\psfrag{Om/Delta}[b][b][1.]{$\Delta \omega/\gamma$}
\psfrag{e=0}[b][b][.9]{$\epsilon=0$}
\psfrag{e=3}[b][b][.9]{$\epsilon=3$}
\psfrag{a}[b][b][.9]{{\bf(a)}}
\psfrag{b}[b][b][.9]{{\bf(b)}}
\psfrag{c}[b][b][.9]{{\bf(c)}}
\psfrag{d}[b][b][.9]{{\bf(d)}}
\centerline{\includegraphics[width=75mm,clip=true]{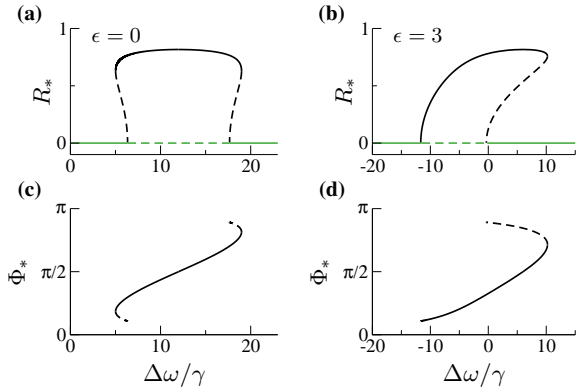}}
\caption{Bifurcation diagrams of synchronized (black) and
incoherent (green) states of Eqs.~\eqref{OA} 
for $K/\gamma=6$, obtained using Eq.~\eqref{bif_d}. 
(a,b) Amplitude $R_*$ and (c,d) Phase difference $\Phi_*$ 
between the Kuramoto order parameters for (a,c) $\epsilon=0$ and (b,d) $\epsilon=3$.}
\label{Fig4}
\end{figure}

%%%%%%%%%%%%%%%%%%%%%%%%%%%%%%%%%%%%%%%%%%%%
A major simplification occurs assuming that $f^\sigma$  evolve in the 
so-called OA manifold, Eq.~\eqref{OAA}, as
the system of Eqs.~\eqref{modes}
becomes independent of the index $l$.
Then, solving the integrals in Eq.~\eqref{op}
by virtue of the residue theorem, 
we find a system of two complex-valued
ordinary differential equations for the
$Z_\sigma(t)=f_1^{\sigma}(\omega=\bar\omega_\sigma-i\gamma,t)^*$
\begin{equation}
\dot Z_\sigma =i \left[\hat{\tilde \omega}_\sigma Z_\sigma - \tfrac{K_{\sigma E}}{2} ( Z_\sigma^2 Z_E^* + Z_E) +\tfrac{K_{\sigma I}}{2} ( Z_\sigma^2 Z_I^* +Z_I)  \right],
\label{OA0}
\end{equation}
with 
$\hat{\tilde \omega}_\sigma \equiv \bar\omega_\sigma +K_{\sigma E} - K_{\sigma I}+i\gamma $.
Restricting our analysis to the case defined by Eqs.~\eqref{K},
Eqs.~\eqref{OA0} reduce to a three dimensional system 
for the amplitudes $R_\sigma$
and the phase difference $\Phi \equiv \Psi_E-\Psi_I$. 
The analysis becomes further facilitated
restricting to the symmetric subspace
\begin{equation}
R_E=R_I \equiv R,
\label{symsol}
\end{equation}
in consistency with our numerical observations,
the transverse stability of the fixed points~
\footnote{See Supplemental Material, Section III.A.},
and related work~\cite{MBS+09}. Hence we analyze the planar system
\begin{subequations}
\label{OA}
\begin{eqnarray}
\dot R &=&R \left[ -\gamma  +\tfrac{K}{2} (1-R^2)  \sin \Phi \right], 
\label{OA_R}\\ 
\dot \Phi &=& \Delta \omega   + K \left[ (1+R^2)  \cos \Phi -2 +\epsilon  (1-R^2) \right].
\label{OA_Phi}
\end{eqnarray}
\end{subequations}
%%%%%%%%%%%%%%%%%%
Besides the fixed point at $R_*=0$, corresponding to incoherence, 
the nontrivial fixed points of Eqs.~\eqref{OA} satisfy 
\footnote{
In contrast with the KM with bimodal frequency distribution
~\cite{Kur84,Cra94,BNS92,MPS06,MBS+09,PM09,PDD16}, our numerical simulations did not reveal
states with time-varying $R_\sigma$. The same occurs in other variants of the 
KM, see e.g.~\cite{HS12}.}
\begin{equation}
\frac{\Delta \omega}{\gamma}= 
 \left[2+ \epsilon(R_*^2-1) \right] \frac{K}{\gamma}  \pm 
  (R_*^2+1)\sqrt{\frac{K^2 }{\gamma^2}-\frac{4}{(1-R_*^2)^2}}. 
\label{bif_d}
\end{equation}   
Figure~\ref{Fig4}(a) displays $R_*$
obtained from Eq.~\eqref{bif_d} for $\epsilon=0$. 
In this case the transitions to synchronization are 
hysteretic and the stable synchronized solution (solid black line) 
exists only in an interval of values of $\Delta\omega>0$. 
As the self-coupling terms are increased, 
Fig.~\ref{Fig4}(b) shows that 
the region of stable synchronization becomes broader, and invades negative values of $\Delta \omega$, see also Figs.~\ref{Fig3}(a)-(d).
Note that the phase difference $\Phi_*$ between $Z_E$ and $Z_I$ 
increases 
monotonically with $\Delta \omega$, see Figs.~\ref{Fig4}(c,d), 
but lies within the interval $(0,\pi)$, 
and thus excitation always precedes inhibition, see also Eq.~\eqref{OA_R}.

 %%%%%%%%%%%%%%%%%%%%%%%%%%%%%%%%%%%%%%%%%%%%%%%%%%%%%%%%%%%%%%%%%%%%%%%%
%%%%%%%%%%%%%%%%%%%%%%%%%%%%%%%%%%%%%%%%%%%%%%%%%%%%%%%%%%%%%%%%%%%%%%%

Differentiating Eq.~\eqref{bif_d} with respect to $R_*^2$ and equating 
the result to zero, allows to analytically obtain the red 
boundaries in Fig.~\ref{Fig3} 
in parametric form (not shown), corresponding to saddle-node bifurcations. 
As $R_* \to 0$, these bifurcations meet the boundaries Eq.~\eqref{Hopf} 
at codimension-2 points where the instabilities change from sub- to 
super-critical. The exact value of the $K$ coordinate is
\begin{equation}
( K/\gamma )^{\pm}_{c2}=\sqrt{\left( 8-2\epsilon^2  
\mp 2 \epsilon \sqrt{8+\epsilon^2}
\right)/(1-\epsilon^2)}.
\label{c2i}
\end{equation}  
Substituting these values into Eq.~\eqref{Hopf} with $D=0$,
we find the location of the codimension-two points 
represented in Fig.~\ref{Fig3}. 

Finally, we have numerically verified that very similar bistability regions appear 
in the phase diagrams for the noisy 
EI-Kuramoto model Eq.~\eqref{EI-KM} with identical 
oscillators ($D>0$, $\gamma=0$). 
In addition, following~\cite{PR99}, we 
found that the codimension-2
points of the noisy EI-KM are 
located at~\footnote{See Supplemental Material, Section IV.}
\begin{equation}
( K/D )^{\pm}_{c2}=\sqrt{\left( 12-2\epsilon^2 
\mp 2 \epsilon \sqrt{24+\epsilon^2}
\right)/(1-\epsilon^2)},
\label{c2ii}
\end{equation} 
which is strikingly similar to Eq.~\eqref{c2i}, 
but here the points lie at slightly larger $K$ values.

%%%%%%%%%%%%%%%%%%%%%%%%%%%%%%%%%%%%%%%%%%%%%%%%%%%%%%%%%%

\emph{Conclusions:} 
Using the averaging approximation we derived a two-population Kuramoto model 
---that we call EI-KM---
from an EI-network of pulse-coupled, Type 1 oscillators.
The resulting EI-KM displays a transition to synchronization 
that has 
the main features of the EI-based
(also known as PING, pyramidal-interneuron gamma) rhythms~\cite{WTK+00,TS09,Wan10,BW12,WC72,BW03,HM03,BK03}:
(i) Oscillations set in exclusively 
due to the cooperative action of both E and I populations.
(ii) Oscillations emerge if
excitatory dynamics is faster than inhibition, irrespective of $\epsilon$. 
(iii) Otherwise, when inhibition is faster than excitation, 
strong enough self-coupling ($\epsilon>1$) is necessary for 
synchrony to occur.
(iv) Excitation always precedes inhibition ($0<\Phi_*<\pi$).
(v) The transition between incoherence and synchronization is 
often 
hysteretic, see e.g.~\cite{HM03}.
While these results have been rigorously demonstrated in the 
EI-KM with Lorentzian heterogeneities (by means of the OA ansatz), 
perturbative and numerical analysis of the
EI-KM with noise reveal the same global picture.  
 
%%%%%%%%%%%%%%%%%%%%%%%%%%%%%%%%%%%%%%%%%%%%%%%%%%%%%%%%%%%%%%%%%%
%%%%%%%%%%%%%%%%%%%%%%%%%%%%%%%%%%%%%%%%%%%%%%%%%%%%%%%%%%%%%%%%%%
\acknowledgments
 
We acknowledge support by MINECO (Spain) under Projects 
No.~FIS2016-74957-P, No.~PSI2016-75688-P 
and No.~PCIN-2015-127. 
We also acknowledge support 
by the European Union's Horizon 2020 research and innovation
programme under the Marie Sk{\l}odowska-Curie grant agreement No.~642563.

%\bibliography{bibliografia}


\end{document}